# scientific reports

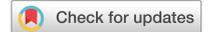

OPEN

# The passband integration properties of Birefringent filter

Xiaofan Wang[1✉], Mikhail Leonidovich Demidov[1,2], Yuanyong Deng[1] & Haiying Zhang[3]

In this article, we discuss an observation phenomenon where the total amount of photons in the full passband of the Birefringent filter is a constant number that is considered by removing the spectrum of the light source irrespective of the instrument transmittance. This conclusion is only noticed and considered to be correct in Huairou Solar Observing Station since 1980's. This article will give a further discussion to the question that had been proposed by the previous researchers. The article structure is organized as history (Sec. 1), experiment (Sec. 2), math (Sec. 3), and discussion (Sec. 4). This issue should be the Paseval-Theorem manifesting itself in astronomical measurement, even though we rigorously demonstrate that this photons conservation has its mathematical generality in Sec. 3.

**BF development in the world.** Lyot-Öhman Birefringent Filter was introduced into solar observation by Öhman and Lyot[1,2]. In 1949, Evans proposed the scheme of the split-element filter which saves one polaroid and this improvement was meaningful since the transmission of polaroid was a problem during those days[3]. The Solc filter invented in 1953 has a slightly narrower, higher transparent profile with higher second side-lobe[4,5]. Evans derived a general expression in 1958 for the transmission of the Solc filter[6]. In 1980, Leroy discussed Solc elements design in detail and provided a summary of principal references about the BF's design and construction from 1933 to 1980[5]. The BF based vector magnetograph was systematically described by Hagyard in 1982[7], especially the very important field of view problem in KD*P (potassium deuterium phosphate/KD$_2$PO$_4$). The off-axis effect of the split-element BF was discussed by Deng in 1997[8]. The design and building of BF had been thoroughly analyzed in a series papers by Title[9–14], and the using of Michelson Interferometer Element as a substitution of narrow Lyot-element was proved to be very successful from GONG on the ground (Global Oscillation Network Group) to the space missions of MDI/SOHO (Michelson Doppler Imager, the Solar & Heliospheric Observatory[15]) and HMI/SDO (the Helioseismic and Magnetic Imager, the Solar Dynamics Observatory[16]).

**BF in China.** BF were developed in China since 1963 when Ai Guoxiang tried to repair a broken BF imported from former Soviet Union in 1958. Since then, the designing and building of BF have never stopped in Chinese solar-physics community. In 1984, a BF based vector magnetograph started its routine observation in HSOS (the first solar station in China, belonging to NAOC). This vector magnetograph is a non-realtime dual channels BF (5324/4861 Å switching). In 2007, a vector magnetograph of three channels BF applying the polarization splitting prisms was finished in HSOS (5250/5247/5173 Å). China has also exported many sets of BFs to other countries, especially by Nanjing Institute of Astronomical Optics & Technology (NIAOT, a branch department of NAOC). In recent years, a new type of BF "8-channels" was put forward for Chinese space solar telescope project (sample machine finished without solar observation test yet[17], shown in Fig. 1). The self-development of solar instruments including the BF is also the major factor that Chinese solar-physics community still holds its relatively strong position in Chinese astronomical community.

**Photon constant problem in BF adjustment.** The BF's solar image quality is mainly determined by the designing and building process. In addition, the adjustment and calibration of BF phase angles for all elements are also an important factor in determining the quality of the passband profile. Both the spectrometer and the photomultiplier tube (PMT) can be used to adjust the BF profile. By using the spectrometer, the spectral resolution should resolve the narrowest BF element. By using PMT[18], the adjustment accuracy can reach to BF's 1/50–1/100 FWHM sampling from heliocentric region on a transparent cloudless noontime (a weighted integration of the solar spectrum, the first of Equation array (2)). Through repeated trials, Hu and Ai realized there was certain constrained relation among the solar originated photons, BF passband, and the scatter photons. It was men-tioned that "The BF passband integration should have constant value no matter what the number of elements or the combination of every element's phase angle is". There were some ambiguities in this Chinese publication[18], especially about the definition of the integration range[18]. But those tiny problems are not important at all when they are compared to the significance of Hu and Ai's pioneering work which provided a framework for subsequent researches. They used Chinese computer TQ-16 (120k calculations per second) to numerically verify this "constant" up to 29 elements' BF, even though it is not yet feasible to build in practicality.

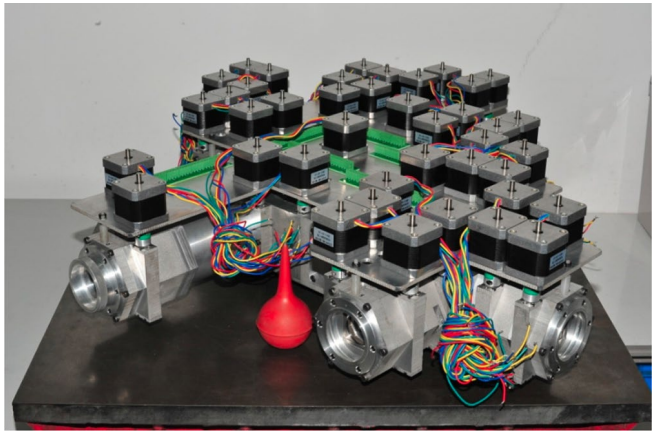

**Figure 1.** The prototype of Chinese 8-Channel BF filters was proposed by HSOS[17] (photographed by Haiying Zhang). The optical part was designed by HSOS, and the mechanical and electrical parts was built by NIAOT (Nanjing Institute of Astronomical Optics & Technology). The Flat fields, polarization modulation, passband stability, temperature control and satellite data bandwidth for this magnetograph will be the main challenges in the future.

We had tried to consolidate their work by testifying this constant properties through the BF scanning "spectrum" realized by rotating wave-plates. But it was difficult due to the earth atmospheric variation and other problems. Until recently, we acquired a very excellent BF wavelength scanning data sample by Liquid Crystal Variation Retard (LCVR)[19]. Hence, we're going to show this properties through observational results in "Solar observation samples and data processing" section. In "Mathematical Generality of Passband Integration" section, we prove the mathematical generality of this BF properties. In "Summary and discussion" section, we discuss the physical meaning and possible future application.

## Solar observation samples and data processing

**General formula of BF.** Different mathematical expressions about BF profile can be found in many literatures. These Four equations in (1) appeared in different publications[2,3,6,20].

$$\tau = \cos^2 \pi n_1 \cdot \cos^2 \pi 2n_1 \cdot \cos^2 \pi 4n_1 \cdots \cos^2 \pi 2^{l-1} n_1 ,$$
$$\tau_L = (t_L/2)[\cos \gamma_1 \cos 2\gamma_1 \cos 4\gamma_1 \cdots \cos 2^{N-1} \gamma_1]^2 ,$$
$$A'^2 = A^2 \cos^2 \frac{\pi \mu e}{\lambda} \cdot \cos^2 \frac{2\pi \mu e}{\lambda} \cdot \cos^2 \frac{4\pi \mu e}{\lambda} \cdots \cos^2 \frac{2^{n-1} \pi \mu e}{\lambda} ,$$
$$I = 0.5 \cdot \cos^2 \frac{\delta_1}{2} \cdot \cos^2 \frac{\delta_2}{2} \cdot \cos^2 \frac{\delta_3}{2} \cdots \cos^2 \frac{\delta_n}{2} ,$$

(1)

where $t_L$ in second equation is a factor representing absorption and interface reflection losses in the filter. The 0.5 factor in the fourth equation, also the denominator of $t_L/2$ in the second equation, represents of inputting the unpolarized light. The first equation neglecting 0.5 factor considers polarized light. A main shortcoming of BF in solar observation is its low throughput comparing to the spectrograph ($\sim 0.5 \cdot t^{(n+1)}$, n is the elements of filter and t is the polarizer transmittance). The interface reflection losses can be largely reduced by using proper silicon oil, but the absorption of crystal and materials is inevitable.

The observed BF profile should be the product of the solar spectrum and the BF passband profile, when sunlight is the source of illumination. For the observation of BF installed inside the telescope without using other dispersion device (such as spectrograph), this product should be an integration within the BF bandwidth (the first formula of eq. array (2)). For the spectrometer observation when the BF has two operations cutting in-and-out between the solar light and the spectrometer, the second formula of eq. array (2) is directly applied ($T_{BF}$=1 for cut-out).

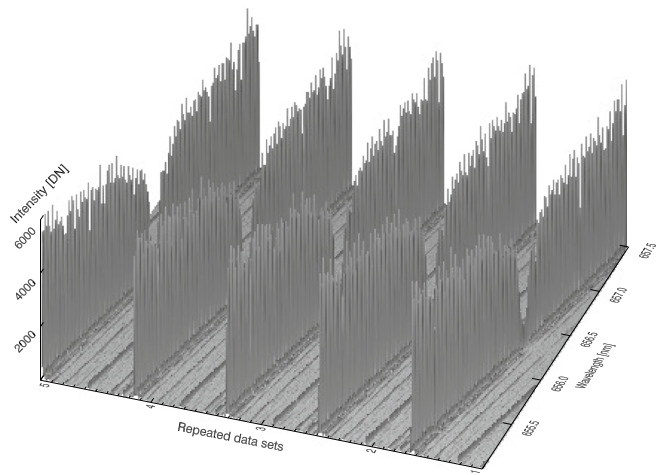

**Figure 2.** Bird's-eye view of five data sets. These data sets including the spectrum were observed by Dr. Hagino and his colleagues[19]. The pixel resolution is 12.087 mÅ along the spectral dispersion direction and the LCVRs scanning step is 123.16 mÅ along the Data sets direction.

$$
\begin{aligned}
I_{Obs}(x, y, t) &= \int_{\lambda_1}^{\lambda_2} I_{Sun}(x, y, \lambda, t) \cdot T_{BF}(\lambda - \lambda_0) d\lambda, \\
I_{Obs}(y, t, \lambda) &= I'_{Sun}(y, t, \lambda) \cdot T_{BF}(\lambda - \lambda_0),
\end{aligned}
\tag{2}
$$

where the variable $\lambda_0$ is the center of BF passband profile, which is controlled by of the wave-plates. In the following discussion, $\lambda_0$ is also the function of time, and time are also corresponded to the LCVR scanning steps.

**Passband shift and data set preview.** The BF passband center can be shifted to different spectral positions with the profile shape almost unchanged. We usually name this operation as "BF passband whole shift" (PWS). In PWS mode, for the type of rotating wave-plate BF the step driver of the thickest element rotates at the maximum angle, and yet the step driver of the thinnest element rotates at the minimum angle.

The scanning spectrum measured by rotating wave-plate PWS mode is difficult to be used for the spectrum study because the scanning time scale is larger than the time scale of solar targets or earth atmospheric fluctuations. It is the similar reason that HSOS applied 3 pieces of DKDP as electro-optic crystal modulator in its first magnetograph, in which 2 modulators work for the velocity field measurement and 1 works for the magnetic field measurement[21].

LCVR seems to be very good electro-optic crystal modulator for BF. Low voltage LCVR is much safer to human or electronic devices than DKDP traditionally used in HSOS, but its nonlinear control model is more complicated than DKDP's linear model, and its anti-electronic jamming is not as good as DKDP. In 2014, Hagino published a set of very excellent solar observations based on a universal LCVRs tunable BF. The multi-variables composite function mapping the voltage to LCVR phase delay can be found in their paper[19]. In this study, we use the same data set to discuss BF photon conservation.

The Fig. 2 represents the bird's-eye view of BF scanning data sets. The LCVRs scanning operation begins from the red wing to the blue wing of the Hα line along abscissa axis with a pixel's spectral resolution 12.09 mÅ, and repeats 5 times along vertical axis with a LCVRs scanning resolution 123.16 mÅ/step (tuning speed > 10 step/second). In these scanning data sets, the highest points in many pulse-like profiles outline 5 Hα profiles which could become continuous and smooth if we could enhance the LCVRs scanning resolution to match the spectral resolution.

The Fig. 3 is a color-scale display to the first data set in Fig. 2. It could also be understood as a view from top to bottom in order to check the symmetry of these profiles. In principle, neglecting the crystal material problem, the function of BF passband should be symmetric with respect to its passband profile center, as long as every element's maximum point are aligned with each other and the spectra of illuminating source are also approximately symmetric to that profile center. The later condition is usually not strictly satisfied because of the properties of the solar spectra and the pre-interference filter. From Fig. 3, it is obvious that the state of BF could be improved better since the trails of its side-lobes are not perfectly symmetric with respect to the central "red-green spine" (also see the red profile in Fig. 5).

Referencing the second formula in Eq. (2), we have two types of spectrum observed by the same spectrograph system. One is the solar spectrum $I'_{Sun}(y, t, \lambda)$ drawn by the red Hα curve in Fig. 4. The other are the overlaid LCVR scanning spectra $I_{Obs}(y, t, \lambda)$ formed by PWS. Those scanning spectra constitute a closed Hα envelope. Every fixed point on BF profile, whatever chosen from main lobe or side lobe, will depict a Hα spectrum in the process PWS (please reference the multiplication operation in the second formula of Eq. (2)). We normalized the maximum of these LCVR scanning spectra to the maximum of the red Hα curve. If we imagine the envelope as a continuous mountain, then the red curve will be its highest spine because of the limitation on transmittance ($T_{BF}(\lambda - \lambda_{PWS}) < 1$). However, there are still several "imperfect" profiles which could be the abrupt increasing of atmospheric transparency or detector problem (Fig. 4).

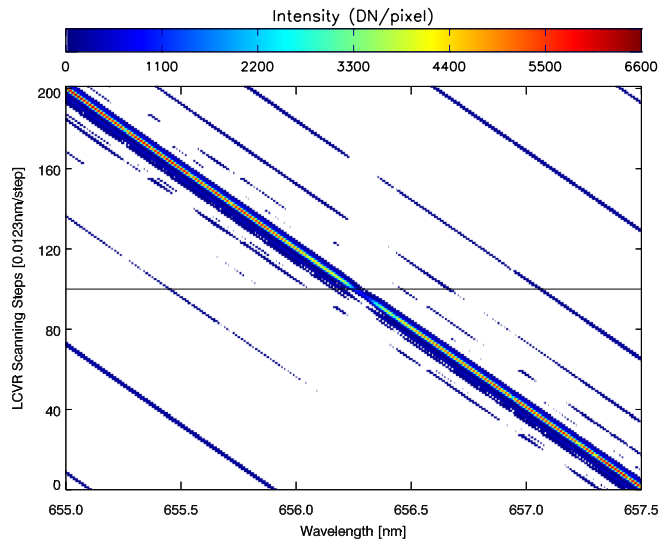

**Figure 3.** The display of first data set of Fig. 2 for checking these profiles' symmetry. The trails of main-lobe and other side-lobes on both sides are clear in the coordinates of scanning step and wavelength. The size of side-lobes and their symmetry are the criterion of the BF adjustment.

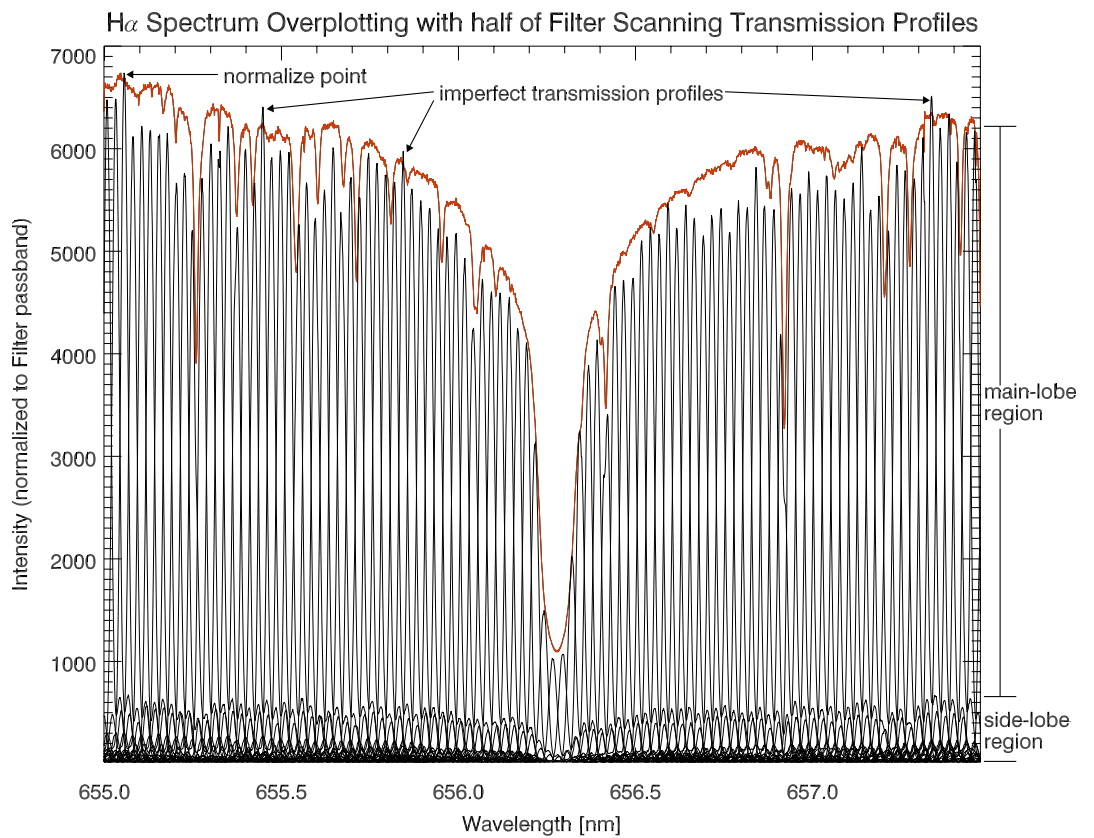

**Figure 4.** The 101 BF profiles half-sampled from the first data set are plotted together with a solar Hα spectrum marked by the red-solid line. All these scanning profiles and this solar spectrum are observed by the same spectrograph system. In a very ideal situation, it could be very close to the red Hα spectrum by joining the highest points of these scanning profiles. The results is the same if this reference point is fixed on any point of these scanning profiles. That is the reason why there are many Hα-like envelopes formed in side-lobes region.

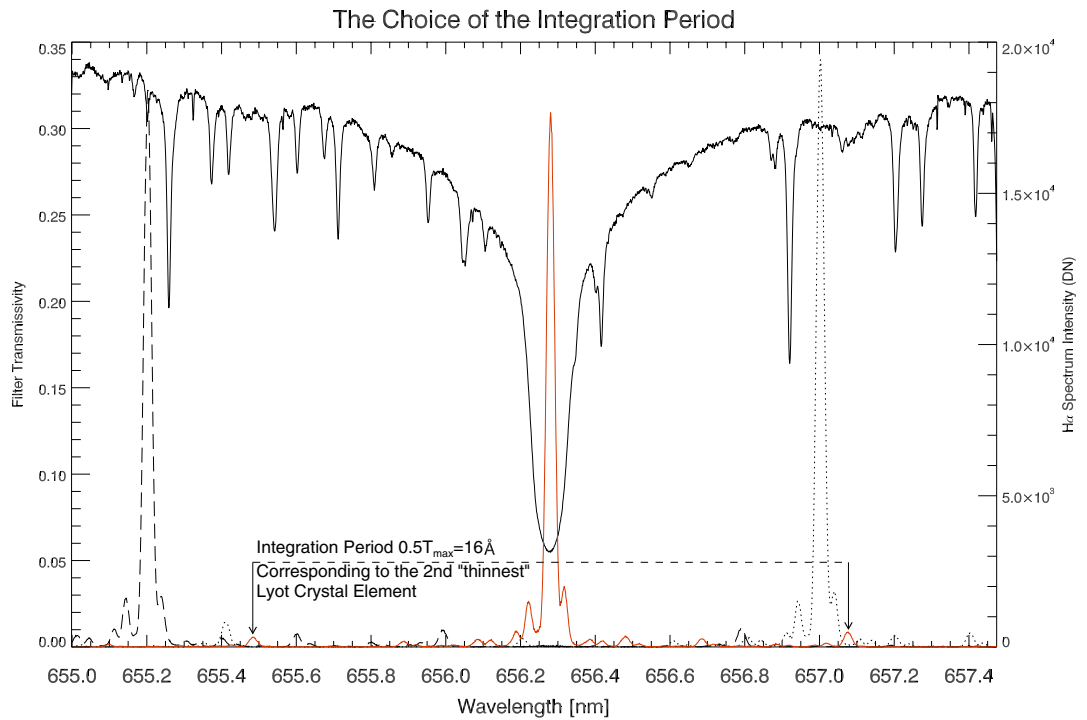

**Figure 5.** The integration range. According to our definition, the integration period should be 32 Å which is out of the range of camera. We have to use two types of integration: one is fixed period marked by the two solid arrows corresponding to the 2nd element period; the other is variable integration range. The BF peak transmittance is about 30% as the same as Fig. 6 in Hagino et al. (2014). The instrument materials absorbtion problem can be attributed to the coefficient $A^2$ in the third formula of equation array (1).

**Integration period and integration constant.** Let us integrate the photons in a range of the BF spectrum. According the second equation of (2), we have the following integration,

$$
\begin{aligned}
C &= \int_{\lambda_1}^{\lambda_2} T_{BF}(\lambda - \lambda_{PWS}) d\lambda \\
&= \int_{\lambda_1}^{\lambda_2} \frac{I_{Obs}}{I'_{Sun}} d\lambda,
\end{aligned}
\tag{3}
$$

in which $I'_{Sun}$ is the solar spectrum in cut-out mode and $I_{Obs}$ is the transmission spectrum of sunlight passing through the BF in cut-in mode. The integration interval in the equation of (3) should be *the full period of the thinnest crystal element which is 32 Å*, but the full camera only covers 24 Å (their experimental device was not built for the purpose of this article). Therefore, we have to smartly choose the fixed integration width and the variable integration width.

For the fixed type, the 16 Å corresponds to the 2nd thinest crystal element and concentrates about 70% of photon energy (see Fig. 5). In each data set, we use a 16 Å sliding-window keeping its center point coinciding with the maximum point of BF passband, and then slide it along the red-green area in Fig. 3 (also the red-spine in Fig. 4). We calculate the integration in every sliding window and repeat this operation for the five data sets. The error bar is defined by 2 times peak-valley value (± PV) calculated from the five data sets. The results marked by the red solid curve on Fig. 6 are very close to the dash-dotted line 0.25. There remains some spectrum residual signal in this red curve. This could be avoided by using stable spectrum-flat light source.

For the variable type, we define an interval variable integration according to the Eq. (3), ranging from 0 Å to 24 Å. In each data set, we select the data from the line cutting through Hα center (see the horizontal solid back line on Fig. 3. Hence, the variable integration range are extended symmetrically from the maximum point of one

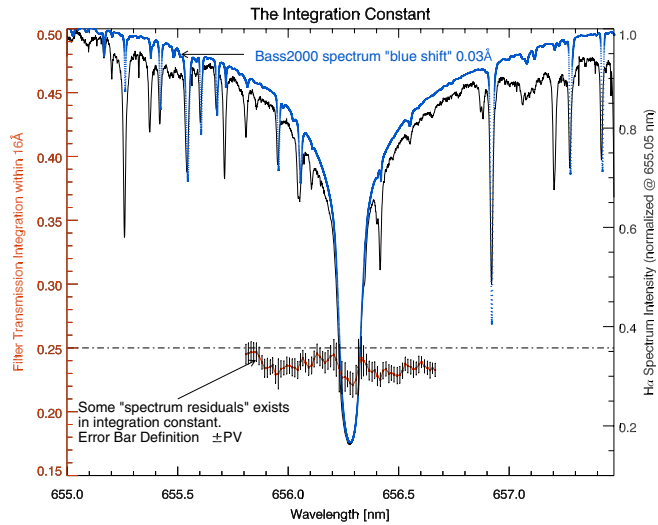

**Figure 6.** The BF Integration Constant. The third formula in equations array (1) and the factor "$A^2 = 0.5$" are contained to calculate the integration within half of this BF widest period. Two comparing solar spectrum normalized at 655.05 nm are plotted together. The blue color Bass2000 Atlas spectrum is shifted a little ($-30$ mÅ) in order to be aligned with the observed solar spectrum (black solid line). The red curve should be close to a "flat" constant level, but some residuals signals exists in the results probably due to the atmospheric fluctuation during the LCVRs scanning.

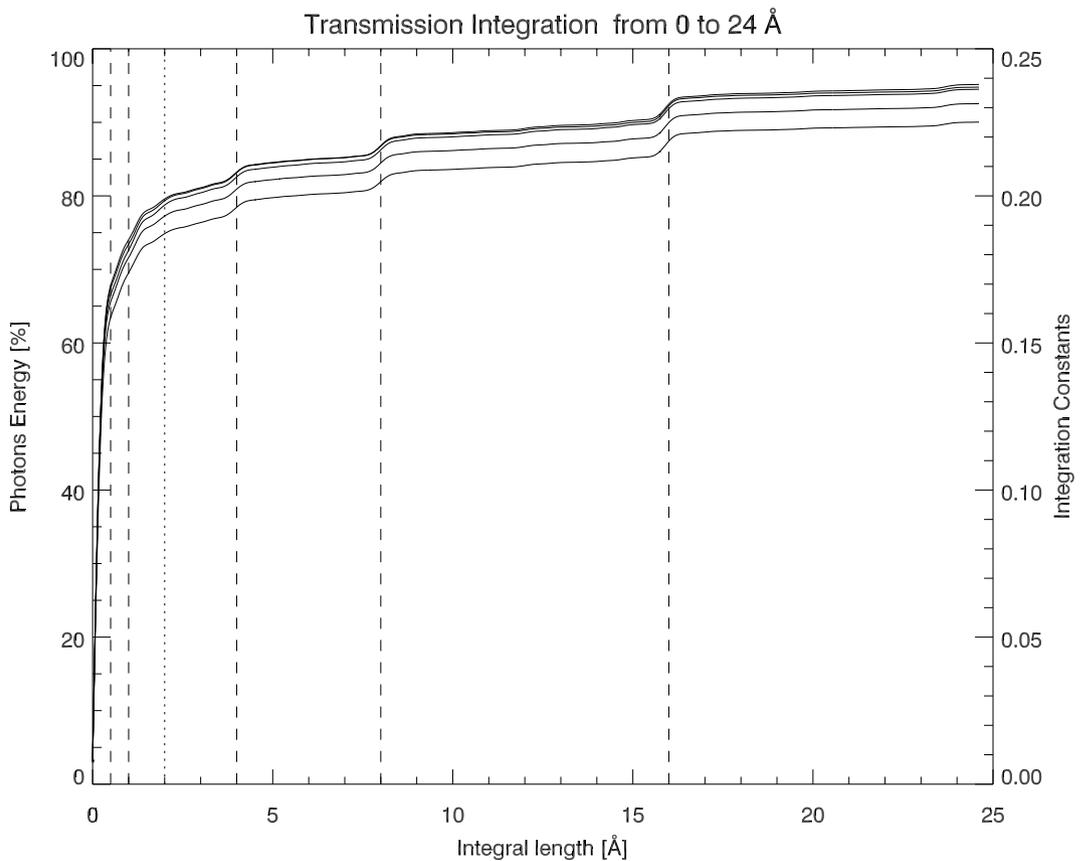

**Figure 7.** Integration variation with different wavelength range. The 80% transmission photons concentrate in the profiles formed by the first three thickest elements. The increase of these curves slows down when the integration range approaches to the 32 Å. The integration value will be invariable on condition that the integration range is fixed on 32 Å, and also the effect of illuminating source should be largely removed from the integration.

| Data-Set 1 | | Data-Set 2 | | Data-Set 3 | | Data-Set 4 | | Data-Set 5 | |
|---|---|---|---|---|---|---|---|---|---|
| Wav. | Integ. | Wav. | Integ. | Wav. | Integ. | Wav. | Integ. | Wav. | Integ. |
| 6563.66 | 0.240 | 6563.68 | 0.238 | 6563.66 | 0.238 | 6563.66 | 0.236 | 6563.66 | 0.233 |
| 6563.54 | 0.247 | 6563.56 | 0.243 | 6563.53 | 0.240 | 6563.53 | 0.241 | 6563.54 | 0.238 |
| 6563.41 | 0.250 | 6563.42 | 0.246 | 6563.39 | 0.246 | 6563.43 | 0.240 | 6563.43 | 0.239 |
| 6563.31 | 0.253 | 6563.29 | 0.248 | 6563.29 | 0.251 | 6563.29 | 0.244 | 6563.29 | 0.237 |
| 6563.18 | 0.251 | 6563.17 | 0.238 | 6563.17 | 0.244 | 6563.17 | 0.235 | 6563.18 | 0.229 |
| 6563.07 | 0.236 | 6563.05 | 0.227 | 6563.05 | 0.231 | 6563.04 | 0.228 | 6563.07 | 0.225 |
| 6562.93 | 0.231 | 6562.93 | 0.224 | 6562.94 | 0.224 | 6562.93 | 0.225 | 6562.93 | 0.221 |
| 6562.81 | 0.230 | 6562.81 | 0.229 | 6562.79 | 0.226 | 6562.81 | 0.228 | 6562.81 | 0.221 |
| 6562.68 | 0.233 | 6562.67 | 0.232 | 6562.68 | 0.228 | 6562.68 | 0.232 | 6562.68 | 0.227 |
| 6562.58 | 0.236 | 6562.55 | 0.235 | 6562.56 | 0.227 | 6562.56 | 0.236 | 6562.56 | 0.225 |
| 6562.43 | 0.234 | 6562.43 | 0.235 | 6562.43 | 0.229 | 6562.43 | 0.235 | 6562.43 | 0.223 |
| 6562.31 | 0.247 | 6562.30 | 0.244 | 6562.30 | 0.235 | 6562.31 | 0.239 | 6562.30 | 0.228 |
| 6562.18 | 0.252 | 6562.18 | 0.251 | 6562.18 | 0.245 | 6562.18 | 0.240 | 6562.19 | 0.239 |
| 6562.07 | 0.253 | 6562.04 | 0.254 | 6562.06 | 0.249 | 6562.03 | 0.247 | 6562.07 | 0.242 |
| 6561.92 | 0.252 | 6561.92 | 0.250 | 6561.92 | 0.248 | 6561.91 | 0.246 | 6561.92 | 0.244 |
| 6561.80 | 0.249 | 6561.80 | 0.247 | 6561.80 | 0.244 | 6561.80 | 0.244 | 6561.80 | 0.241 |
| 6561.67 | 0.247 | 6561.67 | 0.245 | 6561.67 | 0.241 | 6561.68 | 0.241 | 6561.67 | 0.239 |
| 6561.56 | 0.248 | 6561.54 | 0.248 | 6561.56 | 0.244 | 6561.55 | 0.244 | 6561.56 | 0.241 |
| 6561.44 | 0.250 | 6561.43 | 0.250 | 6561.44 | 0.246 | 6561.42 | 0.246 | 6561.43 | 0.243 |
| 6561.31 | 0.254 | 6561.29 | 0.251 | 6561.29 | 0.249 | 6561.29 | 0.248 | 6561.32 | 0.246 |
| 6561.17 | 0.247 | 6561.16 | 0.248 | 6561.17 | 0.244 | 6561.17 | 0.245 | 6561.17 | 0.242 |
| 6561.06 | 0.242 | 6561.05 | 0.243 | 6561.06 | 0.240 | 6561.06 | 0.239 | 6561.08 | 0.236 |

**Table 1.** Test the integral constant at 110 LCVR passband positions from the 5 data sets. Each value (Integ., dimensionless) is integrated in the interval of 20 Å. The center wavelength (Wav., [Å]) of passband corresponding to each integral value is marked in the table.

BF profile. The five curves corresponding to the five data sets in Fig. 7 should approach 0.25 if we integrate them to 32 Å. And the vertical dashed lines in Fig. 7 mark the six BF elements with period 0.5, 1, 2, 4, 8, and 16 Å. The 80% transmission photons are concentrated in the 2 Å period corresponding to the thickest three elements.

Therefore, the results of 0.25 in Figs. 6 and 7 is just a theoretical limit value. In order to give the digital statistical analysis results more intuitively, we calculated the integral of 20 Å for the 110 BF passband positions according to the formula (3). The results are summarized in Table 1. Because of the existence of photons' scatter noise, it is impossible to approach integration constant for a real BF. We prefer the third formula of equation array (1) in which the factor "$A^2 = 0.5$" means half photons left after the first polarizer. For the convenience, we prove the generality of integration constancy in following section without multiplying this factor "$A^2 = 0.5$".

## Mathematical generality of passband integration

**Generality Proof.** Hu and Ai testified their conclusion up to 29 elements' BF by TQ-16 with several phase angle combination in 1984[18]. Now, this type of numerical verification becomes very simple for laptop. From the discussion of "Solar observation samples and data processing" section, we find that it should be correct for a specific instrument that its passband profile integration is a constant. But if this conclusion is enlarged to any BF at any state, does it still hold? Why they proposed this issue at that time? One explanation is they should record too much of PMT data on the paper in the process of adjusting BFs, and they also encountered various solar spectra. This integration invariant properties "on" its largest period is easy to understand for the real BF with "finite" number of elements. But they proposed a more general assumption that this integration should be a constant which is 0.5 under the following conditions:[18]

- the element of BF can be any number,
- its passband can be in any "chaos" state (any $\{\beta_n\}$).

Therefore, in the framework of Hu and Ai, our discussion on the BF's passband integral problem strictly involves three aspects: manufacture, adjustment, and the light source. In the previous observation and data section, both the fixed integral interval and the variable integral interval are calculated and analyzed. In this section, the fixed integration period is applied. Here we clarify that our integration range is **the largest period** decided by the **thinnest BF crystal element**. We will apply the orthogonality of integrated functions and a transformation to discuss the generality of this passband integration, combining the two parts of crystal thickness and phase angle.

The rigorous and concise mathematical demonstration of the conservation of integrated transmission of the tunable Lyot Birefringent Filter can also be realized by expanding the integral into series results and then applying the mathematical induction, which had already been given by Su Dingqiang in 1986[22,23].

Let us take the traditional rotating 1/2 wave-plate as an example (the LCVRs is implemented in another way of voltage and phase delay). To simplify the mathematics, we use the third equation in (1), setting $A^2 = 1$, and introduce the transform $\alpha = 2^{n-1}\pi\mu e/\lambda$ and add initial phase terms $\{\beta_n\}$. In PWS mode, every BF element shifts the same wavelength range, while the rotating angles of the corresponding 1/2 wave-plates obey the $2^n$ proportional series relationship. The series $\beta_n$ in (4) are the phase angles controlled by step driver, and some reference also expressed them as $2\beta_n$. But it does not hinder the following mathematical validity that we treat $\beta_n$ as fixed terms (unspecified phase state, some arbitrary passband profile) and write the term containing crystal element thickness as integral variable $\alpha$.

$$
\begin{aligned}
&T_{all} = T_1 \cdot T_2 \cdot T_3 \cdots T_n, \\
&T_1 = \cos^2(\alpha + \beta_1) \qquad \longrightarrow \pi \\
&T_2 = \cos^2(\alpha/2 + \beta_2) \qquad \longrightarrow 2\pi \\
&T_3 = \cos^2(\alpha/4 + \beta_3) \qquad \longrightarrow 4\pi \\
&\quad\vdots \\
&T_n = \cos^2(\alpha/2^{n-1} + \beta_n) \quad \longrightarrow 2^{n-1}\pi \\
&\alpha = \mu e/\lambda,
\end{aligned}
\tag{4}
$$

where e is the thickness of the crystal element, $\mu$ is difference of refractive index ($n_o - n_e$), and $\pi$ is the narrowest period (the transform $\alpha = 2^{n-1}\pi\mu e/\lambda$ changes the way of mathematical proof, but does not change the final physical conclusion). For any $\{\beta_n\}$, any numbers of Lyot elements (n∈N), and constant $\alpha_0$, it is always correct that

$$
\begin{aligned}
(*) \quad &\frac{1}{\pi}\int_{\alpha_0}^{\alpha_0 + T_{max}} T_1 \cdot T_2 \cdot T_3 \cdots T_n d\alpha = 0.5, \\
&\frac{1}{\pi}\int_{\alpha_0}^{\alpha_0 + 2^{n-1}\pi} T_1 \cdot T_2 \cdot T_3 \cdots T_n d\alpha = 0.5.
\end{aligned}
\tag{5}
$$

The equation (∗) has the physical meaning of integral of energy in full domain, which has relationship with the Parseval's theorem and the Rayleigh's theorem (Bracewell's book[22], page 119, problem 24 on page 252). Parseval's theorem can also be found at https://en.wikipedia.org/wiki/Parseval%27s_theorem.

For the example of (4), the max $T_n$ expressed as $T_{max}$ in (5) equals $2^{n-1}\pi$. Then the Eq. (5) can be expressed as

$$
I_n = \frac{1}{\pi}\int_{\alpha_0}^{\alpha_0 + 2^{n-1}\pi} \prod_{k=1}^{n}\left[\frac{1 + cos(2\alpha/2^{k-1} + 2\beta_k)}{2}\right]d\alpha,
\tag{6}
$$

which is always equal to 0.5 for $n \in N$.

In the following, we will prove that Eq. (6) is always equal to 0.5 for $n \in N$ through logic steps of the mathematical induction.

1° **For** $n = 1$,

$$
\begin{aligned}
I_1 &= \frac{1}{\pi}\int_{\alpha_0}^{\alpha_0 + \pi} \frac{1 + cos(2\alpha + 2\beta_1)}{2}d\alpha \\
&= 0.5.
\end{aligned}
\tag{7}
$$

2° **Set** $I_n = 0.5$, $(n \geqslant 1)$.

$$
\begin{aligned}
I_n &= \frac{1}{2^n\pi}\int_{\alpha_0}^{\alpha_0 + 2^{n-1}\pi} \prod_{k=1}^{n}[1 + cos(2\alpha/2^{k-1} + 2\beta_k)]d\alpha \\
&= 0.5,
\end{aligned}
\tag{8}
$$

where $\{\beta_n\}$ are just constant initial phases for every Lyot element. The continuous product of n terms periodic function is still a periodic function and set $f(\alpha)$ as this periodic function.

$$f(\alpha) = \prod_{k=1}^{n}[1 + cos(2\alpha/2^{k-1} + 2\beta_k)] \tag{9}$$

Hence, according to (4) we have

$$f(\alpha + T) = f(\alpha) \tag{10}$$

(set $T = 2^{n-1}\pi$). Again, Eq. (10) is true for any fixed constant sequences of $\{\beta_n\}$. Restate Eq. (8) as

$$I_n = \frac{1}{2^n\pi}\int_{\alpha_0}^{\alpha_0+T} f(\alpha)d\alpha . \tag{11}$$

According to the properties of periodic function and logic setting $I_n = 0.5$, we acquire that

$$\begin{aligned}
&\int_0^T f(\alpha)d\alpha\\
&= 2^n\pi \cdot I_n\\
&= 2^n\pi \cdot 0.5\\
&= T .
\end{aligned} \tag{12}$$

3° **Prove** $I_{n+1} = 0.5$, $(n \geqslant 1)$!

According to Eq. (8), (9) and (11), we have

$$I_{n+1} = \frac{1}{2^{n+1}\pi}\int_{\alpha_0}^{\alpha_0+2^n\pi} f(\alpha) \cdot [1 + cos(2\alpha/2^n + 2\beta_{n+1})]d\alpha ,$$

by applying $T = 2^{n-1}\pi$,

$$I_{n+1} = \frac{1}{2^{n+1}\pi}\int_{\alpha_0}^{\alpha_0+2T} f(\alpha)d\alpha + \frac{1}{2^{n+1}\pi}\int_{\alpha_0}^{\alpha_0+2T} f(\alpha)\cos(\pi\alpha/T + 2\beta_{n+1})d\alpha ,$$

according to (10), (11), and (12),

$$\begin{aligned}
I_{n+1} =& \frac{2T}{4T} + \frac{1}{2^{n+1}\pi}\int_0^T f(\alpha)\cos(\pi\alpha/T + 2\beta_{n+1})d\alpha\\
&+ \frac{1}{2^{n+1}\pi}\int_T^{2T} f(\alpha)\cos(\pi\alpha/T + 2\beta_{n+1})d\alpha\\
=& 0.5 + \frac{1}{2^{n+1}\pi}\int_0^T f(\alpha)\cos(\pi\alpha/T + 2\beta_{n+1})d\alpha\\
&+ \frac{1}{2^{n+1}\pi}\int_0^T f(\alpha'+T)\cos(\pi(\alpha'+T)/T + 2\beta_{n+1})d\alpha'\\
=& 0.5 + \frac{1}{2^{n+1}\pi}\int_0^T f(\alpha)\cos(\pi\alpha/T + 2\beta_{n+1})d\alpha\\
&+ \frac{1}{2^{n+1}\pi}\int_0^T f(\alpha)\cos(\pi + \pi\alpha/T + 2\beta_{n+1})d\alpha\\
=& 0.5
\end{aligned}$$

which is true for any sequences of $\beta_{n+1}$ since there is no any limitation from $\beta_n$ in the above derivation process.

In summary, by combining 1°, 2°, 3° the Eq. (5) is theoretically correct when the integral interval covers the period of the "thinnest" Lyot filter element. Here, $n \to \infty$ is just proved for theoretical rigorousness. It should be pointed out that we treat the $\alpha$-terms as integral variable and the $\beta$-terms as additives in the above proof. The former is related to building BF and the later is related to adjusting BF. Whatever the combination of $\alpha$-terms and $\beta$-terms is, one important thing keeping the above proof to be correct is that the BF passband function is the continued product of periodic doubling $\cos^2$ or $\sin^2$ function. Proof complete!

**Numerical experiment.** In reality, it is narrow enough for astronomy observation when the Lyot filter elements reaches "$n = 10$" according to our historical experience in HSOS. The discussion about the N-elements filters with arbitrary phase combination is purely mathematical hobby. In this subsection, we provide a piece of numerical code written by IDL to show how the integral of BF passband is a constant.

```
pro simulation_verify

level = 24
Inti = dblarr(level)
another = dblarr(level)
beta = dindgen(160000000.)/160000000.*!dpi

for i = 1, 24 do begin
    alpha = dindgen(160000000.)/160000000.*(2.^(i-1.))*!dpi

    m = 1&n = 1
    for j = 1, i do begin
       m = m*[cos(alpha/(2.^(j-1))+randomu(i+j))]^2.
        n = n*[cos(beta*(2.^(j-1))+randomu(i+0.5*j))]^2.
        if j eq i then print,'j = i',j
    endfor

    Inti[i-1] = 1/!dpi*total(m,/double)*(2.^(i-1))*!dpi/160000000.
    another[i-1] = (2.^(i-1))/!dpi*total(n,/double)*!dpi/160000000.
    print,Inti[i-1],another[i-1]
endfor

help, inti
print, inti
help, another
print, another
end
```

In the previous subsection, we have introduced the transform $\alpha = 2^{n-1}\pi\mu e/\lambda$ to simply the mathematical proof (please compare equation array (4) and (1)). The purpose of this transformation is to change the fractional periods into the integer periods, and to simplify the trigonometric function series. However, the numerical verification is simple for the limited BF elements. The following numerical code contains two kinds of verifications: with and without this transformation. Both of them have numerical random phase terms ($\{\beta_n\}$). The input parameter BF elements "level=24" can be run for most of laptops with 16 GB memory (higher levels needs larger array to subdivide the integral). As we mentioned, they testified up to 29 elements' BF by TQ-16 in 1984 (according to the media reports, some similar but better imported computer at that time even costs China one ton of gold). Their result's accuracy was acceptable, but not as rigorous as the followings.

In summary, we have finished the proof of mathematical generality by deduction method and also finished the verification of limited BF elements by numerical experiment.

## Summary and discussion

As the founder of China's first solar observatory, Hu and Ai proposed the concept of photon conservation within BF passband[18]. But only a few their students knew this conclusion which seems to never appear in the solar-physics community. There are some ambiguities in their article such as the integration interval, integral upper and lower limit, etc. The first author of this article accidentally paid attention to this issue many year ago because Hu and Ai wrote "we feel it is right but we cannot mathematically prove it" in that paper. Now we write this article to discuss this issue again when we encountered very good LCVR based BF scanning spectrum observed by Hagino[19] a few years ago. The contributions of this article are mainly three points: proving the generality of mathematics in this problem; correcting some mistakes introduced by them and clarifying some important details; Using the relatively new LCVR based BF scanning spectrum to clarify this integration conservation property, which should be applicable to evaluate the BF quality and might also be appropriate for future observation. The physical properties of this integral conservation have never been discussed by others. Our article attempts to give the following analysis for the first time.

- We believe that the conservation of integral discussed in this article is corresponded to the Rayleigh's Theorem in Fourier Transform which means the total energy of the two transform domain remains unchanged (Bracewell's book[24], page 119). Until now, Fredga and Högbom (1971)[25] is the only paper mentioned that there is Fourier Transform relation in BF (the distribution function of plate angle differences and the transmission profile). The function determined by eq. (1) must satisfy the existence condition of Fourier transform (finite number of discontinued points of the first kind). The expression of eq. (5) has the strong meaning: the total energy of photons corresponding to the integral has limited value (absolutely integrable function).

- Traditional BF attenuates photons too severely and is difficult to be used in the night time astronomy. However, it is noticed that the BF application to the night time astronomy has been developing in the Australian astronomical community[26–28]. Some amazing results have been achieved by their Filters[28]. The conservation property discussed in this article might be applicable to the night astronomy in the future: the integral constant in spectral domain is a strong constraint to the filter's scanning observation, including dim sources.

Hence, we can draw a conclusion that the photons' integration conservation of BF discussed in this paper should be the Rayleigh's Theorem manifesting itself in astronomical measurement. It could also exist in other imaging type Filters such as Fabry-Perot, Michelson Interferometer, or Fourier Transform Spectrometer, because they also unfold the light periodically. The BF's application to the large aperture telescope is promising in the future, especially for the night time astronomy.

## References

1. Öhman, Y. A new monochromator. *Nature* **141**, 157–158. https://doi.org/10.1038/141157a0 (1938).
2. Lyot, B. Le filtre monochromatique polarisant et ses applications en physique solaire. *Annales dAstrophysique* **7**, 31 (1944).
3. Evans, J. W. The birefringent filter. *J. Opt. Soc. Am. (1917–1983)* **39**, 229 (1949).
4. Šolc, I. Birefringent chain filters. *J. Opt. Soc. Am. (1917–1983)* **55**, 621 (1965).
5. Leroy, J. L. Solc elements in Lyot–Öhman filters. *J. Opt.* **11**, 293–304. https://doi.org/10.1088/0150-536X/11/5/002 (1980).
6. Evans, J. W. Solc birefringent filter. *J. Opt. Soc. Am. (1917–1983)* **48**, 142 (1958).
7. Hagyard, M. J., Cumings, N. P., West, E. A. & Smith, J. E. The msfc vector magnetograph. *Sol. Phys.* **80**, 33–51. https://doi.org/10.1007/BF00153422 (1982).
8. Deng, Y., Ai, G. & Wang, J. Inhomogeneous distribution of brightness in the split-element filter. *Appl. Opt.* **36**, 1576–1579 (1997).
9. Title, A. M. Improvement of birefringent filters. *Sol. Phys.* **33**, 521–523. https://doi.org/10.1007/BF00152437 (1973).
10. Title, A. M. Improvement of birefringent filters 2: Achromatic waveplates. *Appl. Opt.* **14**, 229–237 (1975).
11. Title, A. M. Improvement in birefringent filters 3: Effect of errors on wide field elements. *Appl. Opt.* **14**, 445–449 (1975).
12. Title, A. M. Improvement in birefringent filters 4: The alternate partial polarizer filter. *Appl. Opt.* **15**, 2871–2879 (1976).
13. Title, A. M. & Rosenberg, W. J. Improvements in birefringent filters 5: Field of view effects. *Appl. Opt.* **18**, 3443–3456 (1979).
14. Title, A. M. & Ramsey, H. E. Improvements in birefringent filters 6: Analog birefringent elements. *Appl. Opt.* **19**, 2046–2058 (1980).
15. Scherrer, P. H. *et al.* The solar oscillations investigation—Michelson doppler imager. *Sol. Phys.* **162**, 129–188. https://doi.org/10.1007/BF00733429 (1995).
16. Schou, J. *et al.* Design and ground calibration of the helioseismic and magnetic imager (HMI) instrument on the solar dynamics observatory (SDO). *Sol. Phys.* **275**, 229–259. https://doi.org/10.1007/s11207-011-9842-2 (2012).
17. Deng, Y. & Zhang, H. Progress in space solar telescope. *Sci. China Phys. Mech. Astron.* **52**, 1655–1659. https://doi.org/10.1007/s11433-009-0255-2 (2009).
18. Hu, Y. F. & Ai, G. X. The integrated transmission conservation of the tunable Lyot birefringent filter and its adjustment method. *Acta Astrophysica Sinica* **4**, 234–242 (1984).
19. Hagino, M., Ichimoto, K., Kimura, G., Nakatani, Y., Kawate, T., Shinoda, K., Suematsu, Y., Hara, H., & Shimizu, T. Development of a universal tunable birefringent filter for future solar observations. In: *Advances in Optical and Mechanical Technologies for Telescopes and Instrumentation, Proc. SPIE* **9151**, 91515V (2014). https://doi.org/10.1117/12.2055728.
20. Huang, Y. R., Xu, A. A., Tang, Y. H., Xuan, H. C. & Qin, Z. H. *Observational Astrophysics* 421st edn, Vol. 7 (China Science Press, Beijing, 1987).
21. Ai, G. *et al.* The birefringent filter for measuring solar vector magnetic field and sight-line velocity field. *Sci. Sin. Ser. Math. Phys. Tech. Sci.* **27**, 1086–1095 (1984).
22. Su D. Q. A rigorous demonstration of Integrated transmission conservation of the tunable Lyot Birefringent Filter and a method to measure the integral light amounts outside the mainband. *Acta Astrophysica Sinica* **6**, 72 (1986).
23. Su D. Q. A rigorous demonstration of the conservation of Integrated transmission of the tunable Lyot Filter. *Chin. Astron. Astrophys.* **10**, 81 (1986)
24. Bracewell, R. N. *The Fourier Transform and Its Applications, 3rd edn*. McGraw-Hill Education (Asia) Co. and China Machine Press, Beijing (2002).
25. Fredga, K. & Högbom, J. A. A versatile birefringent filter. *Sol. Phys.* **20**, 204–227. https://doi.org/10.1007/BF00146111 (1971).
26. Giovanelli, R. G. & Jefferies, J. T. On the optical properties of components for birefringent filters. *Aust. J. Phys.* **7**, 254. https://doi.org/10.1071/PH540254 (1954).
27. Bland-Hawthorn, J., van Breugel, W., Gillingham, P. R., Baldry, I. K. & Jones, D. H. A tunable Lyot filter at prime focus: a method for tracing supercluster scales at z ~ 1. *Astrophys. J.* **563**, 611–628. https://doi.org/10.1086/323770 (2001).
28. Bland-Hawthorn, J. & Kedziora-Chudczer, L. Taurus tunable filter—7 years of observing. *Publ. Astron. Soc. Aust.* **20**, 242–251. https://doi.org/10.1071/AS02023 (2003).

## Acknowledgements

The authors are grateful to Dr. Hagino and his colleagues for building this BF as state-of-the-art and providing very excellent LCVRs scanning data. Wang Xiaofan would like to extend his appreciation to Dr. Wang Dongguang for her description about HSOS earlier history, to Dr. Lin Jiaben for his comment about DKDP, and to Dr. Hou Junfeng for his discussion about the transmissivity of polarizer and wave-plate. This work and the authors have been supported by the following projects and grants: Natural Science Foundation of China (Project No. 11573042, 11103041, U1331113, 11427901, 11427803, U1531247); CAS President's International Fellowship Initiative (PIFI, Project No. 2017VMA0009), Basic Research Program II.16. The author Wang Xiaofan is also partly supported by the Strategic Priority Research Program on Space Science, the Chinese Academy of Sciences (Grant No. XDA15320302, XDA15052200, XDA15320102).

## Author contributions

X.W. prepared the main parts of manuscript. The Filter development history of HSOS was provided by Y.D. M.L.D. made a very positive contribution to the problem of the Filter's transmission. The Figure 1 was photographed and provided by H.Z. All authors reviewed the manuscript.

## Competing interests

The authors declare no competing interests.

## Additional information

**Correspondence** and requests for materials should be addressed to X.W.

**Reprints and permissions information** is available at www.nature.com/reprints.

**Publisher's note** Springer Nature remains neutral with regard to jurisdictional claims in published maps and institutional affiliations.

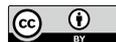